\documentclass{emulateapj}
\usepackage{natbib}
\shorttitle{Tidal stirring of satellites with shallow density profiles prevents them from being too big to fail}
\shortauthors{Tomozeiu et al.}

\begin{document}

\title{Tidal stirring of satellites with shallow density profiles prevents them from being too big to fail}

\author{Mihai Tomozeiu, Lucio Mayer}
\affil{Institute for Computational Science, University of Zurich, Winterthurerstrasse 190, CH-8057 Z\"{u}rich, Switzerland}
\affil{Physics Institute, University of Zurich, Winterthurerstrasse 190, CH-8057 Z\"{u}rich, Switzerland}
\email{mihai@physik.uzh.ch}
\and
\author{Thomas Quinn}
\affil{Astronomy Department, University of Washington, Box 351580, Seattle, WA 98195, USA}


\begin{abstract}

The ``too big to fail'' problem is revisited by studying the tidal evolution of populations of dwarf satellites with different 
density profiles. The high resolution cosmological $\rm \Lambda CDM$ "ErisMod" set of simulations is used. These simulations can model 
both the stellar
and dark matter components of the satellites, and their evolution under the action of the tides of a MW-sized host
halo at a force resolution better than 10 pc. The stronger tidal mass loss and re-shaping of the mass distribution
induced in satellites with $\gamma=0.6$ dark matter density distributions, as those resulting from the effect of feedback in
hydrodynamical simulations of dwarf galaxy formation, is sufficient  to bring the circular velocity profiles in 
agreement with the kinematics of MW's dSphs. In contrast, in simulations in which the 
satellites retain cusps at $z=0$ there are several "massive
failures" with circular velocities in excess of the observational constraints.  
Various sources of deviations
in the conventionally adopted relation between the circular velocity at the half 
light radius and the one dimensional line-of-sight
velocity dispersions are found. 
Such deviations are caused by the response of circular velocity profiles to tidal effects, 
which also varies depending on the
initially assumed inner density profile, and by the complexity of the stellar 
kinematics, which include residual
rotation and anisotropy. In addition
tidal effects naturally induce large deviations in the stellar mass-halo mass relation for halo masses 
below $\rm 10^9 ~ M_{\odot}$,
preventing any reliable application of the abundance matching technique to dwarf
galaxy satellites.

\end{abstract}



\section{Introduction}
The ''too big to fail'' (TBTF) problem is one of the few remaining open issues faced by the 
otherwise very successful $\rm \Lambda CDM$ cosmology.
Essentially a legacy of the missing satellites problem, it has been first highlighted by
\cite{2011MNRAS.415L..40B, 2012MNRAS.422.1203B}, who noticed
that the satellites of the MW-sized halos in the Aquarius 
simulations \citep{2008MNRAS.391.1685S} contained up to ten  
massive objects with circular velocities significantly exceeding those of the
MW dwarf spheroidal (dSph) satellites, yielding correspondingly higher central densities.
When using the abundance matching technique \citep{2004ApJ...609...35K, 2010MNRAS.404.1111G, 2010ApJ...710..903M, 2013ApJ...770...57B}
in order to indirectly estimate the luminosity of the satellites, the authors found 
that such massive satellites would have to be
more luminous than the most luminous dSph satellites of the MW and be comparable with the Magellanic Clouds (MCs).

A flurry of papers has been published with the purpose of investigating this problem 
further \citep{2014MNRAS.444..222G, 2014MNRAS.440.3511T,2015MNRAS.451.2524M,
2015MNRAS.453.3575J}, or to attempt solving it  
\citep{2014JCAP...12..051D, 2015MNRAS.446.2363O, 2015MNRAS.453...29E, 
2016MNRAS.457L..74D, 2015MNRAS.454.2092O, 2015MNRAS.454.2981C, 2015arXiv150804143R, 2014arXiv1412.2748S, 2015MNRAS.450.3920B, 2012MNRAS.424.2715W, 2012ApJ...761...71Z}. 
Conventional solutions proposed generically for the missing satellites problem, such as 
suppression of gas accretion and photoevaporation by the cosmic ionizing background 
\citep{1996MNRAS.278L..49Q,2000ApJ...539..517B,1999ApJ...523...54B} and ram pressure stripping
combined with tidal mass loss \cite{2007Natur.445..738M} would all be weakly effective
or entirely negligible at the mass scale of interest (corresponding to $\rm v_{circ}^{max}$ in the range $\rm 30-70~km~s^{-1}$
).

Supernovae feedback in bursting dwarf galaxies
has been shown to solve the cusp-core issue naturally \citep{2015arXiv150804143R, 2010Natur.463..203G, 
2015MNRAS.454.2981C, 2005MNRAS.356..107R},
and has been proposed also as a way to alleviate the TBTF problem, at least
in the sense that it drives a decrease of the central density. \cite{2014ApJ...786...87B} 
used cosmological hydrodynamical simulations in which satellites
of a MW-like galaxy appear to overcome the TBTF problem via core formation.  \cite{2016MNRAS.457.1931S} argue that baryonic feedback effects in satellites inhabiting a Local Group-like cosmological hydro simulation are sufficient to avoid the TBTF problem without core formation. However, force resolutions in current cosmological hydro simulations of
MW-like galaxies are at best
a few hundred pc, namely comparable with the effective radius of some MW dSphs.
This makes it numerically problematic to study robustly
the evolution of the baryonic and dark matter components of satellites under the action of the host tidal field \citep{2004ApJ...608..663K}.  
Therefore at the moment a generally accepted and numerically robust solution is still lacking \citep{2015PNAS..11212249W}.

In the article by \cite{2016ApJ...818..193T} we have shown, using a cosmological simulation with satellites
resolved to better than 10 pc force resolution, that the tidal evolution of satellites with
shallow density profiles as those resulting from the supernovae-driven outflows is markedly
different from that of cuspy CDM satellites.  In particular, we showed that one of the proposed scenarios for the origin of the MW dSphs, tidal stirring of disky satellites resembling present-day dIrrs \citep{2001ApJ...559..754M},
is much more efficient in satellites with shallow profiles. Since the TBTF problem is borne out of a comparison
with the kinematics of the MW dSphs it is natural to ask if our findings could have implications on the TBTF problem.  
Here we use a more extended set of simulations to show
that the different tidal evolution of cuspy and shallow satellites provides a natural solution to the
TBTF problem.

Our new analysis builds on 
work done by other authors on the subject 
\citep{2014JCAP...12..051D, 2015MNRAS.446.2363O, 2015MNRAS.453...29E, 2015arXiv150804143R},  
but is unique because of the numerical technique and resolution achieved.
We use cosmological N-body simulation of MW-sized halos 
which can follow both the stellar components and the dark matter components
of satellites. They were run using the new massively parallel ChaNGa N-Body+SPH code 
\citep{changa2010, 2015ComAC...2....1M}. Using the
replacement technique \citep{2005MNRAS.364..607M} applied to the ErisDark
simulation \citep{2014ApJ...784..161P}, we can follow
the stellar component of satellites rather than just the dark matter as they
evolve in the tidal field of the cosmological primary host halo, and can do that 
at very high stellar mass and force resolution (0.3 $\rm M_{\odot}$ and 4 pc, respectively). 
The cosmological simulations are briefly described in section 2.

In section 3 we present the results derived from the ErisMod simulations 
that are relevant for the TBTF problem topic. Finally, in section 4 we discuss the 
implication of the results and we list the conclusions.

\section{Numerical simulations and data}

All the results presented in the letter are obtained from analysing the last two ErisMod pairs of simulations - 
ErisMod High and ErisMod Low (paper in preparation). 
The initial conditions of the simulations are based on the dark matter only zoom-in cosmological simulation 
ErisDark run \citep{2014ApJ...784..161P}  which is 
faithful to the $\Lambda$CDM cosmological model($\rm H_0$ = 73 km $\rm s^{-1} Mpc^{-1}$, $\rm n_s = 0.96$,$ \rm ~\sigma_8 = 0.76$, 
$\rm \Omega_m = 0.268$ and $\rm  \Omega_{\lambda} = 0.732$).
In each of the simulations a MW-size halo of mass $\rm 9\cdot10^{11}~M_{\odot}$ assembles by redshift zero inside 
the highly resolved region of one Mpc radius.
The virial quantities of the ErisMod High simulation pair at redshift zero were in both cases
$\rm M_{vir}~ = ~ 9.0\cdot10^{11}~M_{\odot}$ and $\rm R_{vir}~ = ~ 247~kpc$.
The same quantities for the ErisMod Low simulation pair were $\rm M_{vir}~ = ~ 9.1\cdot10^{11}~M_{\odot}$ 
and $\rm R_{vir}~ = ~ 247~kpc$ for both simulations. 

The previously mentioned simulation pairs are generated using the replacement method \citep{2005MNRAS.364..607M}. 
At different epochs of interest a simulation is being stopped and dark matter halos of virial masses in the range $\rm 10^{8.5}~M_{\odot}$ and $\rm 10^{10}~M_{\odot}$ 
found in the proximity and outside of the main halo are being replaced with constructed high resolution models of disk galaxies 
embedded in dark matter halos. The replacing objects were created with the GalactICs code \citep{2008ApJ...679.1239W, 2005ApJ...631..838W, 1995MNRAS.277.1341K} 
and contain one million stars and one million dark matter particles, each. The previously mentioned code generates spherically symmmetric multi-component equilibrium
models with a spherically symmetric total density distribution and an exponential stellar disk with a cuttoff. 

The masses and the softening lengths of the high resolution objects can be as low as 0.3 $\rm M_{\odot}$ and 4 pc, respectively. 

Each of the simulation pairs includes one run in which all halos of interest are replaced by dwarf galaxy halos with a classical 
NFW profile (cusp coefficient - $\rm \gamma~=~1.0$ - Navarro, Frenk, White profile) and one run in which the replacing halos have shallower 
profiles (described $\rm \gamma~=~0.6$). All other free halo parameters - virial mass, virial radius, scale radius - 
as well as all the stellar disk parameters are common to both runs in each simulation pair.   

The virial mass and the virial radius, as defined by \cite{1998ApJ...495...80B}, of a replacing model galaxy match the replaced halo. 
The mass of the stellar disk was set to be three orders of magnitude smaller than the mass of the entire halo. This choice was based on estimates from abundance matching 
\citep{2013ApJ...770...57B, 2014ApJ...792...99S}. All other parameters required for 
the construction of the high resolution models are dependent on the virial mass and virial radius through relations that have been described in the paper by
\cite{2016ApJ...818..193T}. The concentration of each halo is given by the empirical results published by \cite{2011ApJ...740..102K} and is dependent on
$\rm M_{vir}$ and redshift. Moreover the total mass inside the virial volume of an object is by construct equal to the virial mass of the orginal halo.
The value of the scale heights for the stellar distribution is set to be 80 pc for all generated disk galaxies. The scale radius of the exponential disk is
modelled by the equation:

\begin{equation}
R_d = 0.4  \left( \frac{M_{vir}}{10^9 M_{\odot}}\right)^{0.5} kpc.
\end{equation}

The resulting values of $\rm v_{rot}/\sigma_{\star}$ at $\rm r_{1/2}$ are between 1.5 and 2.5 at the time of replacement.  
The details and motivations behind the construction of the multi-component dwarf galaxy models are described extensively in the article by \cite{2016ApJ...818..193T}.

The first set of replacements in the ErisMod High simulation pair was made at redshift 2.9 and the last at redshift 1.6.  
In total there were 4 replacement epochs. Between redshift 2.87 and 2.00 the ErisMod simulations 
presented in the publication by \cite{2016ApJ...818..193T} and ErisMod High simulations are the same runs. Beyond redshift 2.0  
they were run separately. After the last replacement the number of high resolution objects in each of the ErisMod 
High runs was 24.

In the case of ErisMod Low simulation pair the first replacement was made at redshift 1.1 and the last at redshift 0.1.  
In total there were 4 replacement epochs. The final count of high resolution objects in each of the ErisMod Low runs was 17.
 
The progenitors as well as other halos found in the proximity of the main halo at
an epoch of replacement went through a process of selection. 
If the respective dark matter cloud was found to be monolithic, not elongated, not part of a group of overlapping 
halos and with no identifiable satellites, then a high resolution model will be generated and used as replacement. 
This procedure leads to the exclusions of between 40\% and 60\% of all considered halos at a certain epoch of replacement. 
Thus, the objects that are replaced are in general more resilient to the tides due to their compactness.   

The two simulation pairs ErisMod High and ErisMod Low should be seen as complementary since the first epoch of replacement 
for ErisMod Low starts after the last epoch of replacement in the ErisMod High simulation. 

Information regarding the orbits, the original virial mass as well as the redshift of the first pericenter 
passage for all replaced objects is available in figure \ref{fig:figure1}.

\section{Results}

\begin{figure*}
\centering
\includegraphics[width=16cm]{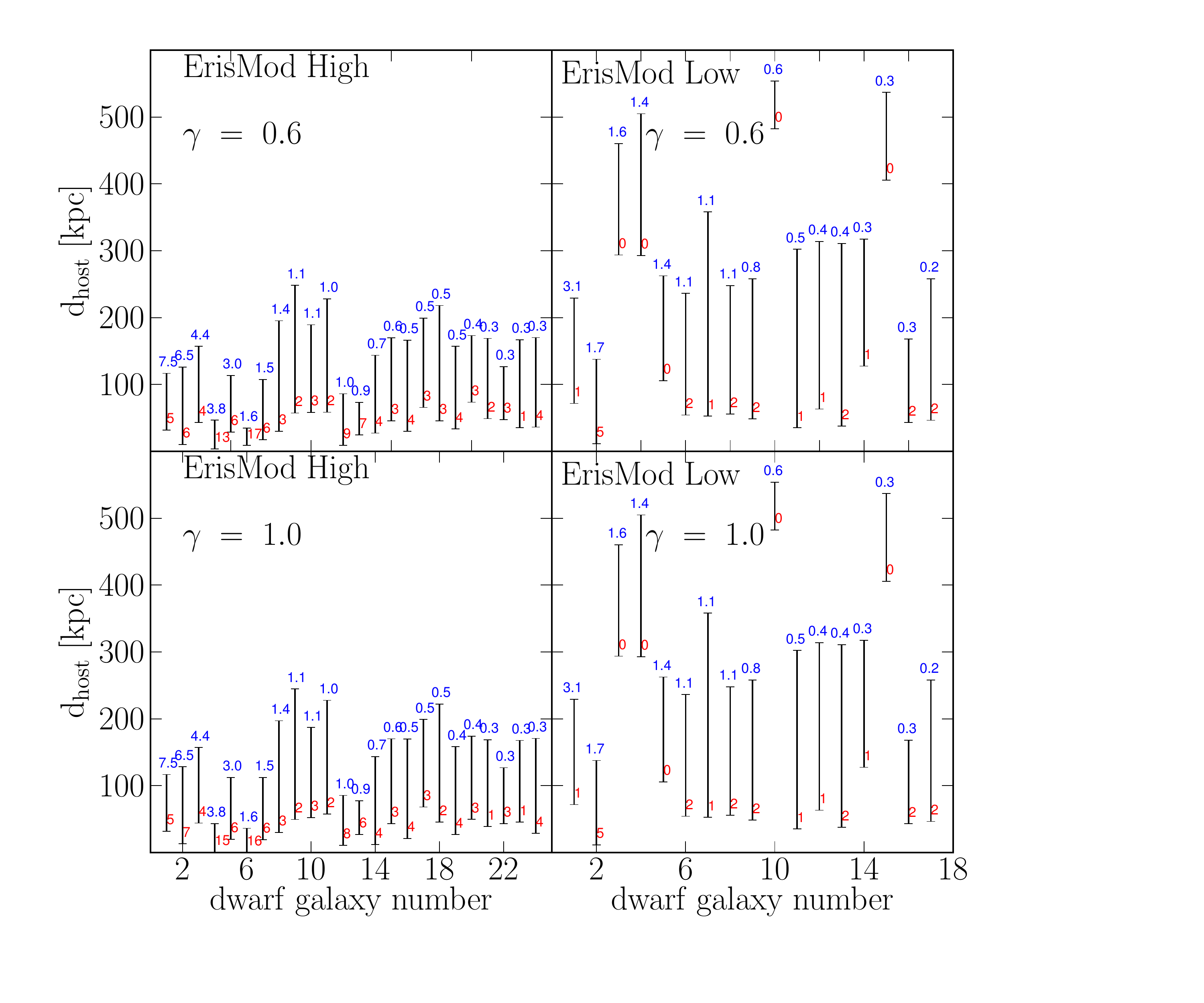}
\caption{\label{fig:figure1}The dwarf galaxy distance from host ranges for the four ErisMod 
simulations are plotted as bars. The dwarf galaxy number is sorted by halo mass at infall. At the top of each bar the marked number (blue color) denotes the pre-infall 
virial mass of the objects in units of $\rm 10^9~M_{\odot}$. Near the bottom of each bar the neighbouring 
number (red color) indicates the count of the pericenter passages done by the galaxy until redshift 0.}
\end{figure*}

For the following analysis we consider subsets of the satellite populations presented in figure \ref{fig:figure1}. 
We are interested in objects that at $\rm z=0$ are within 247 kpc of the centre of the main halo, 
namely inside the virial volume
of the host halo. This was done in order to achieve consistency with previous analysis of the TBTF problem from the literature.
Moreover we exclude 
from analysis all the objects that have lost more than 99\% of their mass before redshift zero, 
as the low number of particles 
will make the results of the analysis not trustworthy. 
This class of objects includes galaxies that have lost such a large fraction 
of their mass that they become irrelevant for the TBTF problem.

The circular velocity of the satellites is defined as the function of distance from centre:
\begin{equation}
v_{circ}(r) = \sqrt{\frac{G M(<r)}{r}}.
\end{equation}

\begin{figure*}
\centering
\includegraphics[width=16cm]{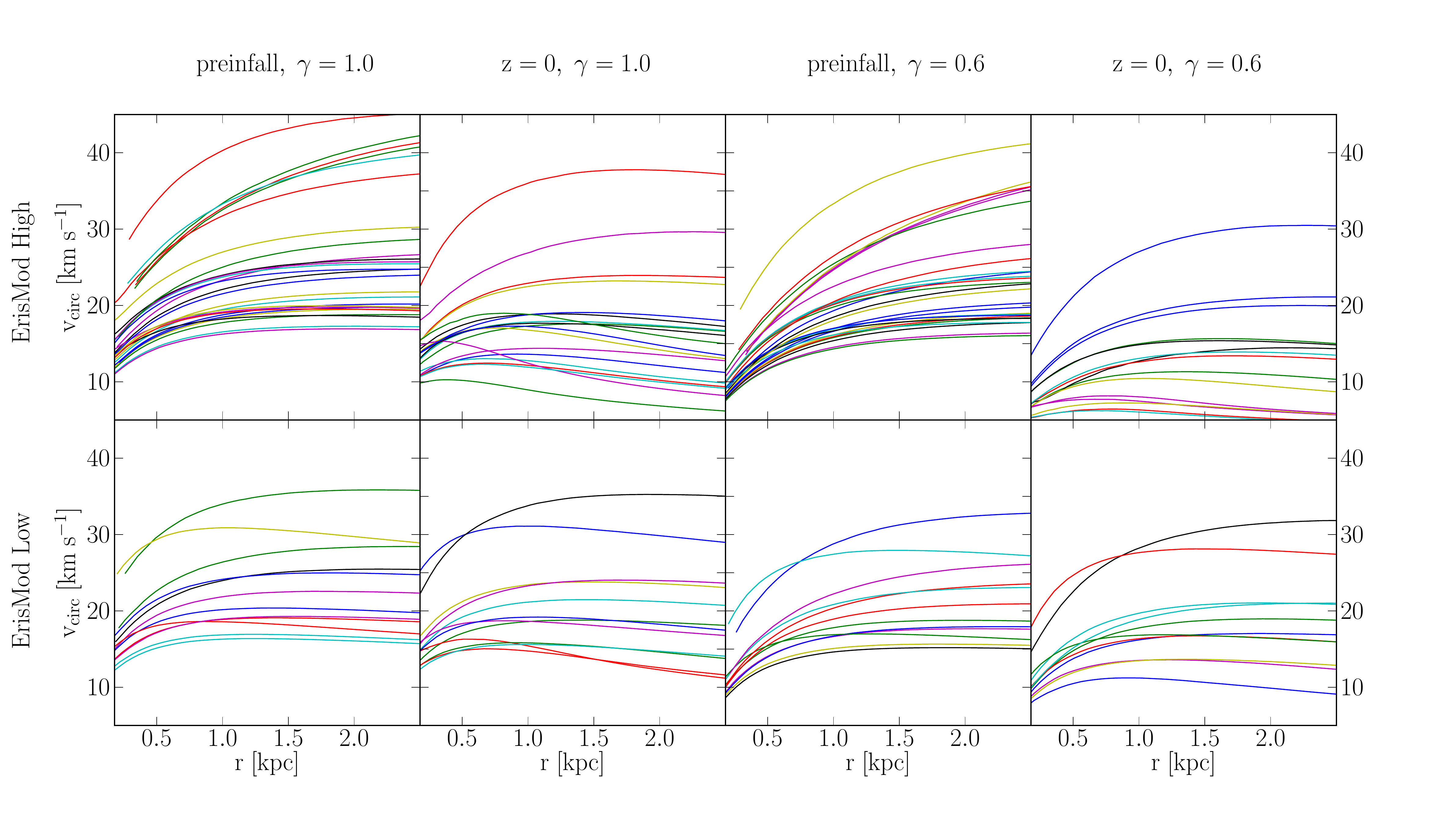}
\caption{\label{fig:figure2} The simulated dwarf galaxies' circular velocity profiles. 
Top/bottom row presents the data from the ErisMod High/Low simulation pairs. The first and the third column 
contain the circular velocity profiles of the of halos with steep ($\gamma = 1.0$) and 
shallow ($\gamma = 0.6$) dark matter distribution prior to infall. The second and the 
fourth column contain the circular velocity profiles of the halos (with steep and 
shallow dark matter distribution) at redshift zero. The maximum circular velocities for most
objects before infall are reached at radii greater than the plots display 
(see figure 5 of \cite{2016ApJ...818..193T}
for more details).}
\end{figure*}

The results 
are shown in figures \ref{fig:figure1} and \ref{fig:figure2}. In figure \ref{fig:figure1} the profiles 
of the halos at the epoch of 
replacement are plotted for comparison.  
Moreover in figure \ref{fig:figure2} we have added the estimated circular velocities at the half light radius
available for the MW dwarf spheroidal satellites based on the data compiled by \cite{2012AJ....144....4M}. 
They are derived from the line of sight velocity dispersion data using the equation:
\begin{equation}
v_{circ}(r_{1/2})=\sqrt{3} \sigma_{\star}.
\end{equation} 
originally proposed by \cite{1999ApJ...522...82K} and later used by \cite{2010MNRAS.406.1220W} and 
in papers on TBTF \citep{2011MNRAS.415L..40B, 2012MNRAS.422.1203B}.

According to figure \ref{fig:figure3} in both simulation pairs the runs in which the original halos were 
replaced with models with shallow dark matter density profiles produce a population of satellites that obey 
the constrains from observations at redshift zero. Note that although there are three simulated 
galaxies with maximum circular velocities above the values given by the observations, at small radii ($< 600$ pc),
which is the region directly probed by the kinematics of the MW dSphs, they
are consistent with the available data.

At the same time the simulations where the initial halos were 
replaced with models with steep dark matter profiles are not consistent with observations as they predict several 
massive satellites that would have velocities above the constraints at all radii. Even adding the
two Magellanic Clouds (not marked in figure \ref{fig:figure3}), would not eliminate the discrepancy.

In figure \ref{fig:figure4} the maximum circular velocities were plotted against the stellar and total bound 
mass. According to the results displayed the amount of stellar mass maintained by a satellite does not have 
a very strong relation, unlike the total mass,  with the value of the associated maximum circular velocity. 
Satellites with similar maximum circular velocities (within a 10\% range) can have stellar masses that vary by 
more than one order of magnitude. Moreover objects with similar amounts of stellar mass 
(within a 20\% range around $\rm 10^{5.5}~M_{\odot}$) can have maximum circular velocities ranging from 7 to 20 $\rm km~s^{-1}$. 
At the same time, the spread in the maximum circular velocity and total bound mass data is much smaller. 

In the characteristic velocities panel of figure \ref{fig:figure5} for around half of the mock observations 
the circular velocities at half light radius estimated with the line-of-sight velocity dispersion are good estimates for the
maximum circular velocities while for the other half they are not. In the characteristic radii panel it can be 
seen that for the majority of the satellites the 3D half light radius is by a factor of 2 or 3 smaller 
than the radius at which the maximum circular velocity is reached. The half-light radii were estimated using 
the three dimensional definition. The values of the radii determined with the 
3D definition are usually 1.3 times larger \citep{2010MNRAS.406.1220W} than the radii obtained with the projection definition 
\citep{ 2012AJ....144....4M},  such as those presented in figure \ref{fig:figure3}.

\section{Discussion and Conclusion}

\begin{figure*}
\centering
\includegraphics[width=16cm]{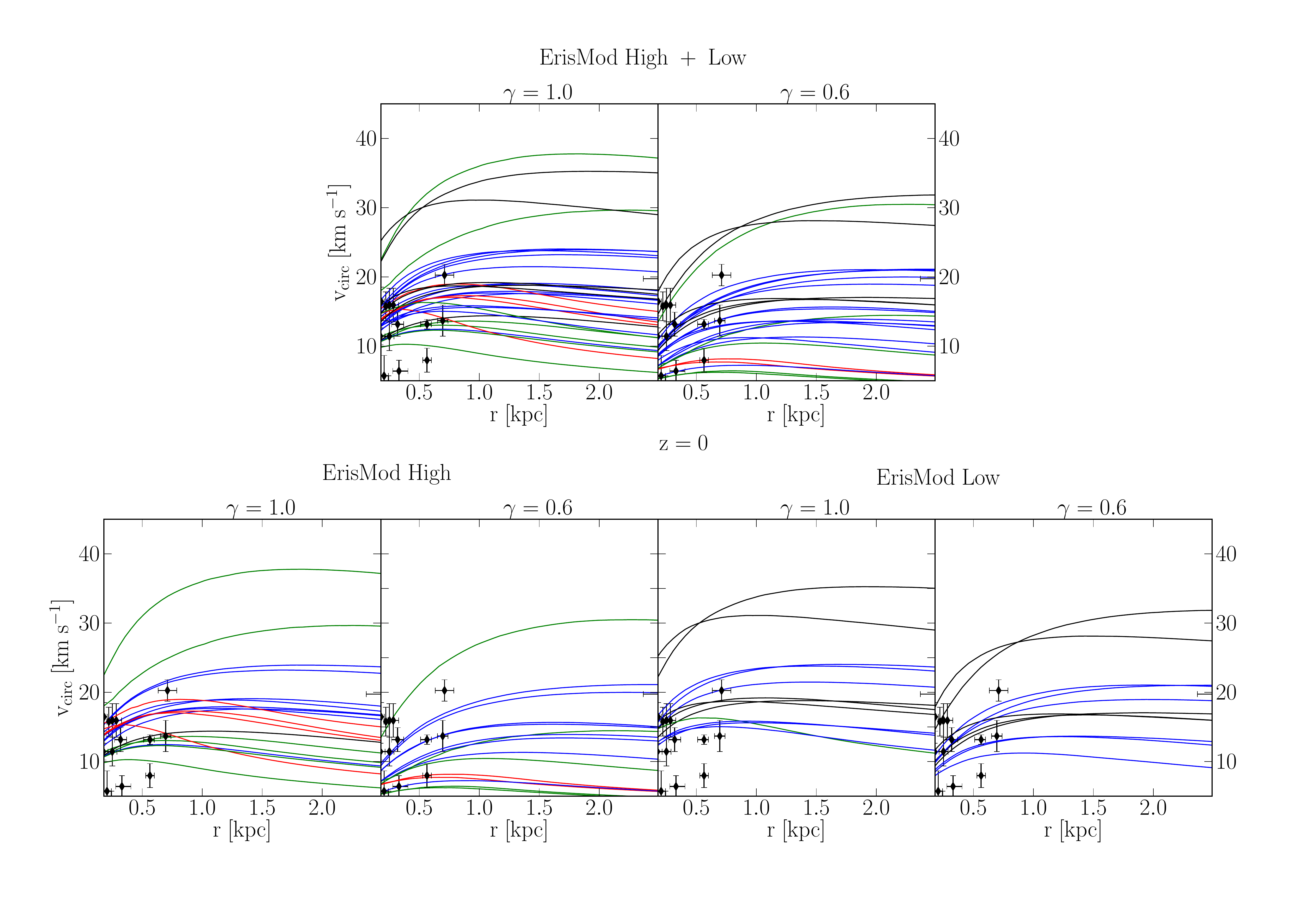}
\caption{\label{fig:figure3} The simulated dwarf galaxies' circular velocity profiles at redshift 0.  
The color scheme marks the number of pericenter passages as follows: red curves correspond to objects with 6 or more pericenter 
passages, green for 4 or 5 pericenter passages, blue for 2 or 3 and black for 1 or no pericenter passages. 
Constraints on the circular velocities and half-light radii of observed MW satellites were marked with black markers 
accompanied by error bars. The satellites are: Bootes I, Draco, Sculptor, Sextans I, Ursa Major I, Carina, Hercules, Fornax, Leo IV,
Canes Venatici I, Leo I. The data for the Magellanic Clouds was not displayed.}
\end{figure*}

The results of section 3 show that the central dark matter density distributions of dwarf galaxy satellites 
before infall play 
a crucial role in solving the TBTF problem. As can be seen in figure \ref{fig:figure3} (top row) 
the population of simulated objects 
with a cusp coefficient $\rm \gamma~ =~ 0.6$ in both ErisMod Low and ErisMod High simulations at redshift 0 are 
consistent with 
present-day observation of the MW dSph satellites. Although the maximum circular velocity for three simulated objects 
is greater than the observed values, the circular velocities of these objects at small radii (200 - 300 pc) are consistent 
with observations in that region. 
At the same time, the population of simulated 
halos with pure NFW profiles 
are not consistent with observations for any radius, with 4 being
above the observational data constraints.
Such a discrepancy would get even worse had we included infalling halos
that could not be replaced for the reasons described in section 2.

According to the first panel of figure \ref{fig:figure5} for a significant fraction of satellites (about a third) the assumptions
$\rm v_{circ}^{r_{max}} \approx v_{circ}^{r_{1/2}} \approx \sqrt{3} \sigma_{\star}$  is not a good approximation. First, for some objects with
many orbits ($>$4) the circular velocity at $\rm r_{1/2}$ can be as low as half the maximum circular velocity (third panel in figure \ref{fig:figure5}).
Moreover, in the second panel, we see that due to two reasons the approximation $\rm v_{circ}^{r_{1/2}} \approx \sqrt{3}\sigma_{\star}$ is
misleading. First there is a population spread (many points reside far from the locus of the $\sqrt{3}/3$ slope). Second, for the dwarf
galaxies that
have retained significant rotation
the direction of observation has a strong impact on the measured velocity dispersion. For the respective
set of galaxies the maximum circular velocity can be  severely underestimated. A line of sight velocity dispersion of $\rm 7~km~s^{-1}$
can correspond to circular velocities in the range 10 to 35 $\rm km~s^{-1}$. Residual rotation occurs more often for the remnants
of cuspy satellites, which transform less efficiently into dSphs as shown in a previous work \citep{2016ApJ...818..193T}.

\begin{figure*}
\centering
\includegraphics[width=8cm]{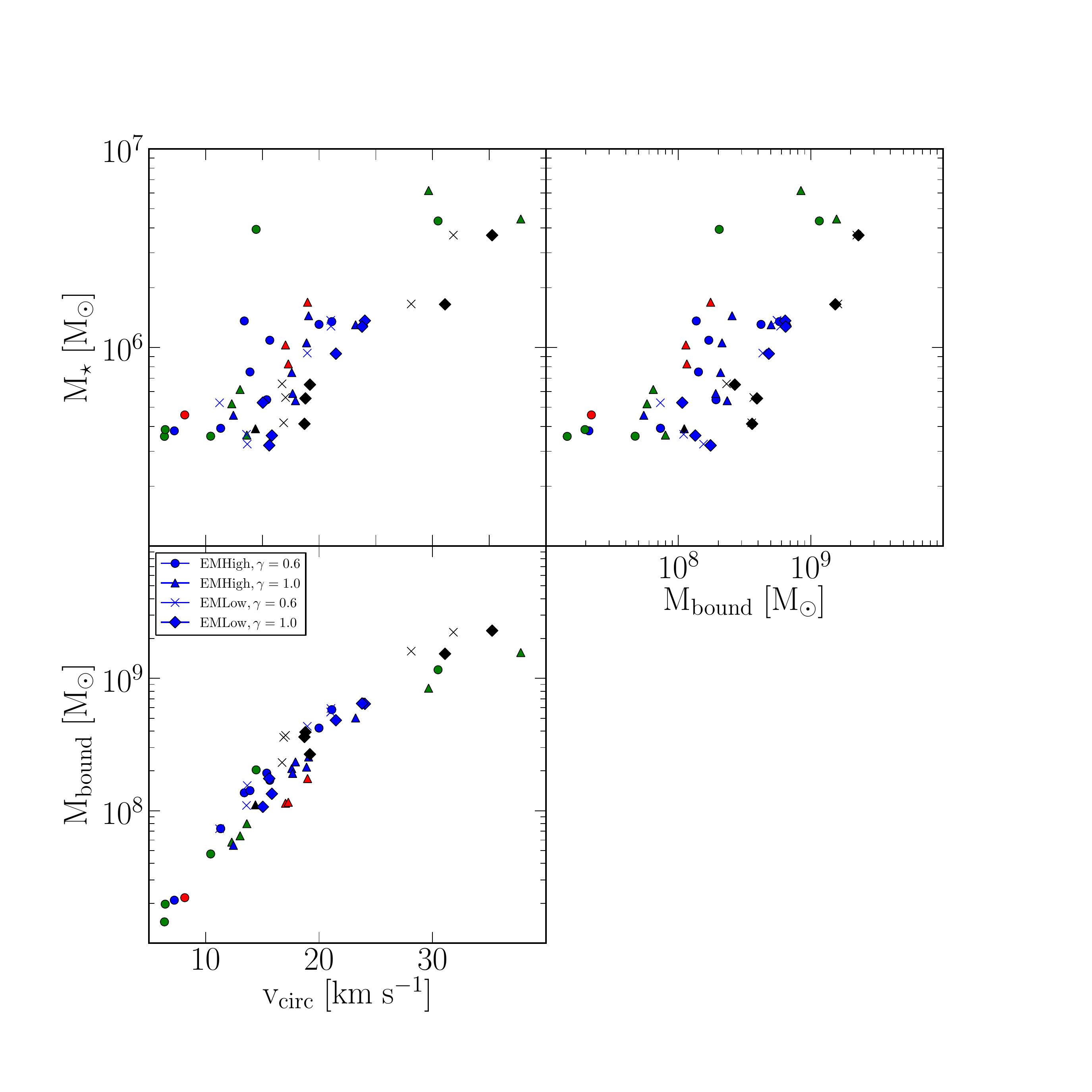}
\caption{\label{fig:figure4}The maximum circular velocity as a function of the bound stellar mass (top left panel) 
and total bound mass (bottom panel). In the left panel the stellar mass was plotted against the total bound mass.
The marker shape differentiate between the different simulations 
and the colors encode the number of pericenter passages that the associated satellite made. The color 
scheme is explained in the caption of figure \ref{fig:figure3}.}
\end{figure*}

\begin{figure*}
\centering
\includegraphics[width=8cm]{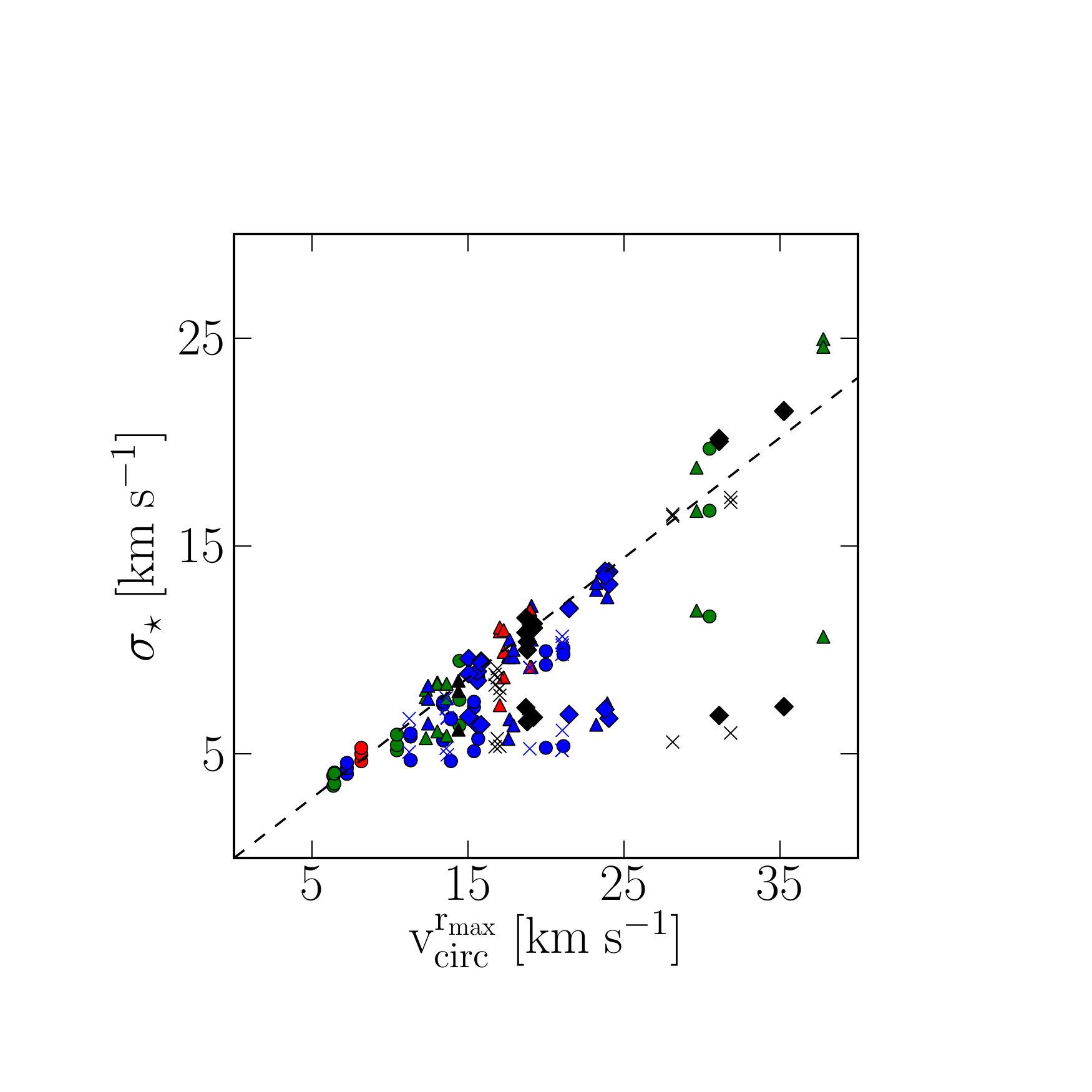}
\includegraphics[width=8cm]{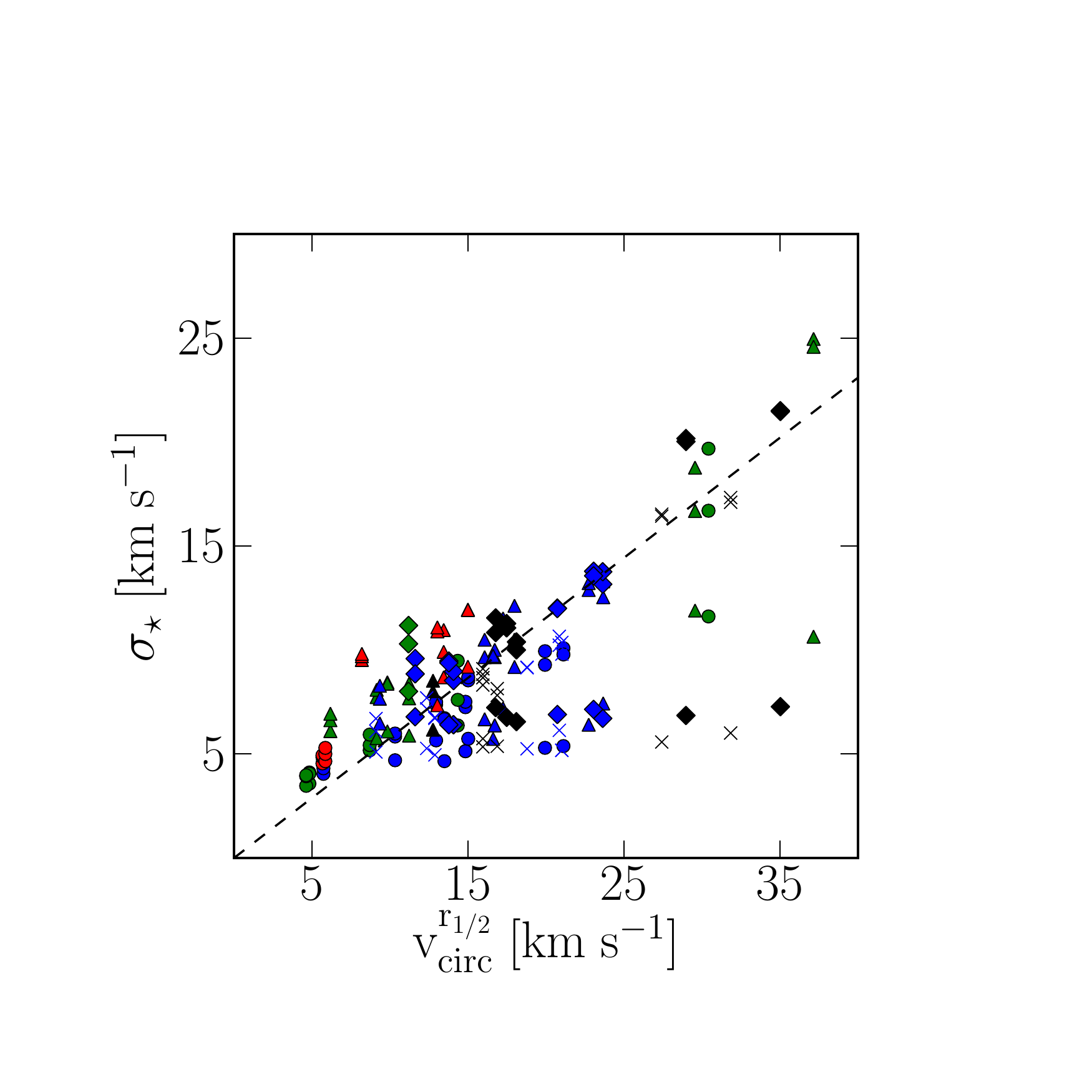}
\includegraphics[width=8cm]{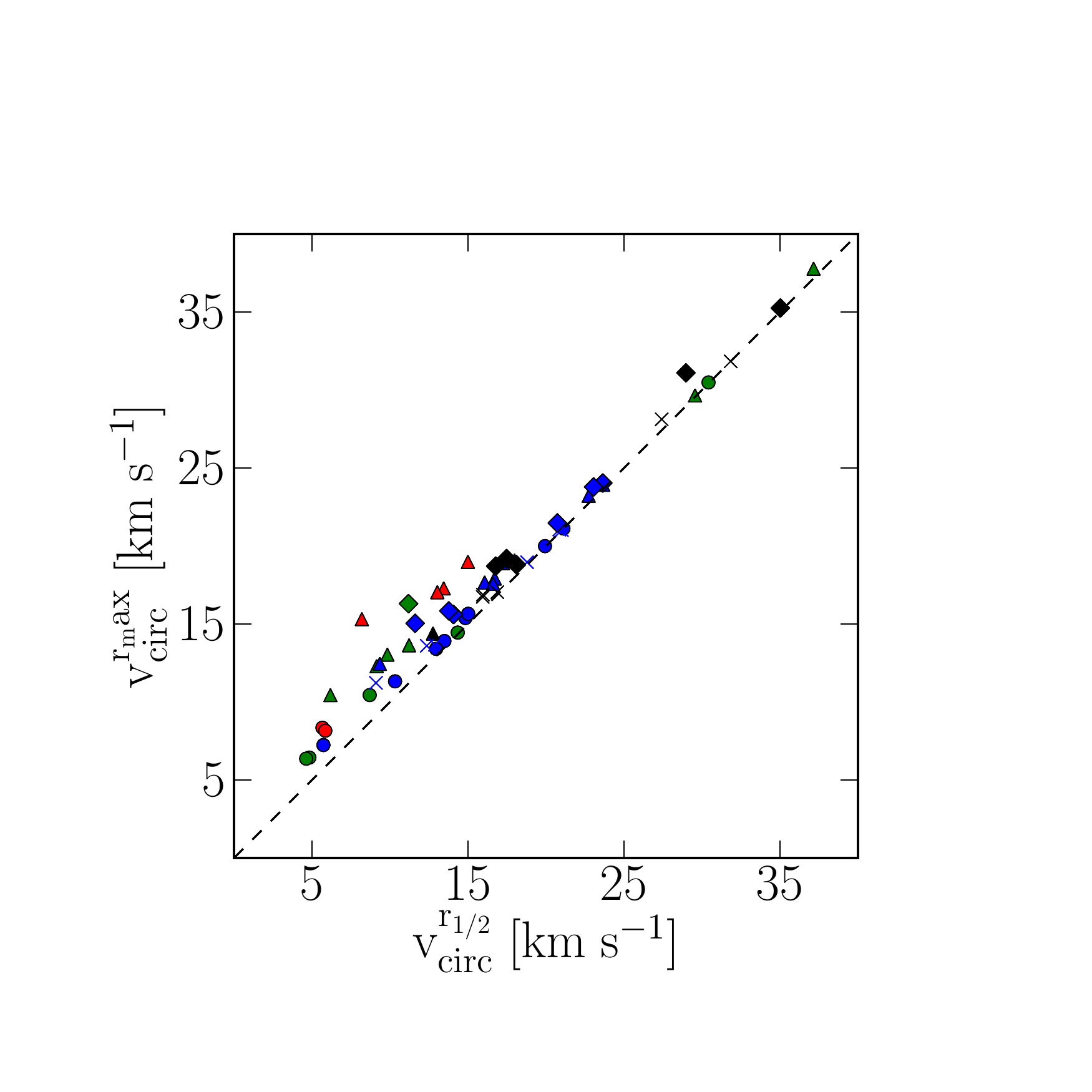}
\includegraphics[width=8cm]{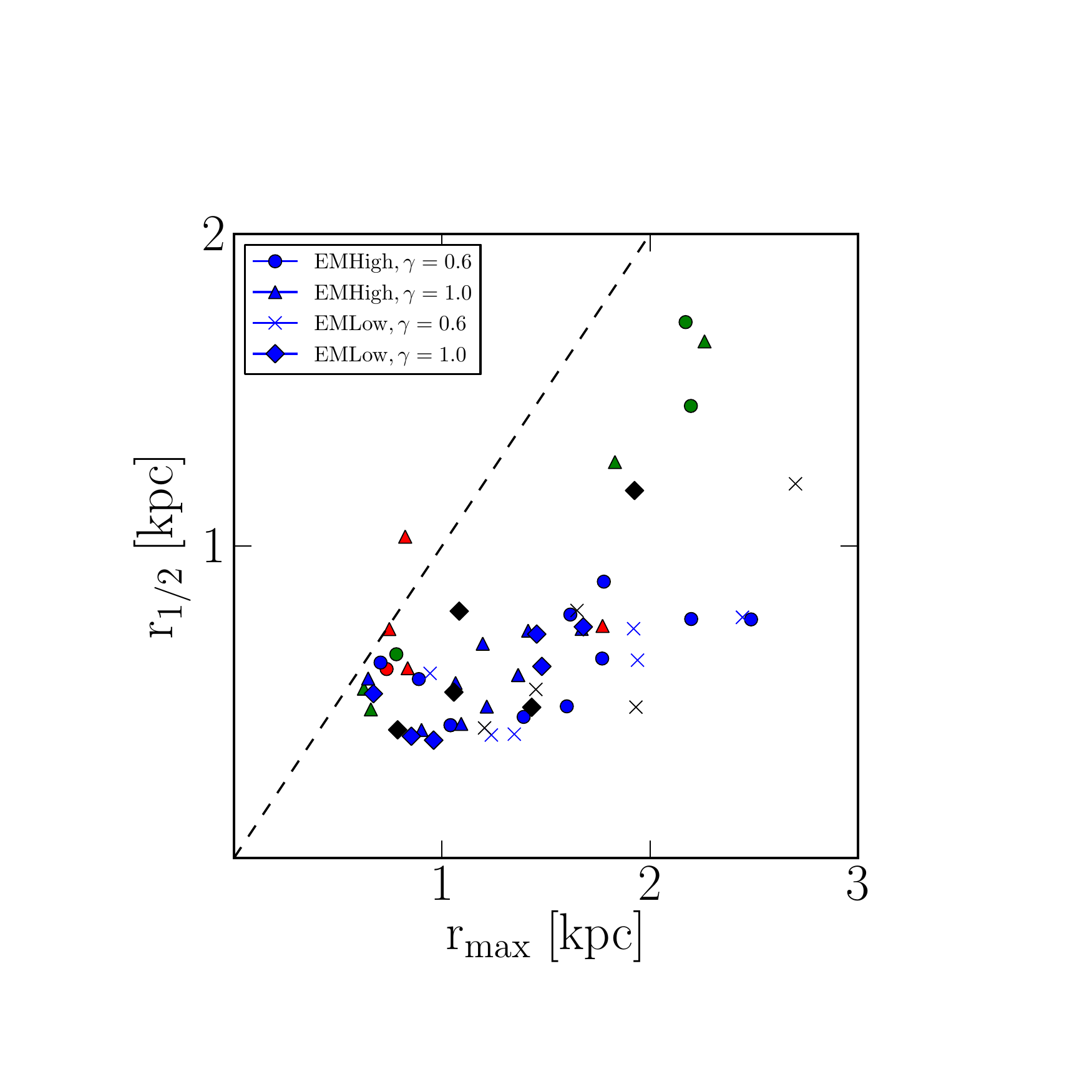}
\caption{\label{fig:figure5}Characteristic velocities and radii for the simulated satellites population. In the upper panels the maximum circular velocities (left panel) and
circular velocity at $r_{1/2} $ were plotted against the simulated line of sight velocity dispersions. 
For each circular velocity three velocity dispersions have been plotted corresponding to 
the direction of the principal axes of the respective object. They are equivalent to the line of sight velocity dispersions if the observer was situated alon a principal axis.
The black dashed line is described by the equation $\rm v_{circ}^{r_{max}/r_{1/2}} ~=~ \sqrt{3}~ \sigma_{\star}$.
In the left bottom panel the maximum circular velocities against the circular velocity at  $\rm r_{1/2} $ were plot. The dashed line corresponds in this case to unity.
In the right bottom panel the radii where the circular velocities reach maximum were plotted against the half light radius.  The black dashed line in the last panel 
mark the place where $\rm r_{1/2}~ = ~r_{max}$. The color scheme and marker shapes are the same as in figures 3 and 4.}
\end{figure*}

We have also found that the abundance matching assumption is not valid for populations of tidally perturbed dwarf satellite galaxies. 
Although prior to infall all objects obey by construct the abundance matching assumption (\cite{2016ApJ...818..193T}, Tomozeiu et al 
in preparation) 
tidal stirring will cause objects with the same circular velocity to span even more than an order of magnitude in total stellar mass 
(figure \ref{fig:figure4}), preventing any simple monotonic relation between stellar mass and halo mass. This is in contrast
with the arguments adopted in early papers on the TBTF problem \citep{2012MNRAS.422.1203B}.

In conclusion we find that the TBTF problem can be solved naturally by combining two general mechanisms in hierarchical galaxy formation.
The first one is the ability of stellar feedback to modify the inner dark matter distribution of low mass halos, which are more susceptible
to outflows, lowering the central density by erasing the original CDM cusp. 
The second one is the natural occurrence of tidal effects after infall, in the form of tidal
mass loss but also repeated tidal shocks that modify the mass distribution of baryons and dark matter further.
While recent hydrodynamical cosmological simulations \citep{2013ApJ...765...22B} have suggested that satellites
with shallow profiles could be the key in solving the TBTF problem,
they have been lacking the sufficient resolution to address the specific effect of tides on such satellites directly, which is an 
advantage of the numerical
technique employed here.
Future hydrodynamical cosmological simulations with comparable resolution will still be needed to 
better quantify our
findings in the complete galaxy formation framework.

\def\apj{ApJ}
\def\apjl{ApJL}
\def\aj{AJ}
\def\mnras{MNRAS}
\def\aap{A\&A}
\def\nat{nature}
\def\araa{ARAA}
\def\pasa{PASA}
\bibliography{ads}

\begin{thebibliography}{}
\expandafter\ifx\csname natexlab\endcsname\relax\def\natexlab#1{#1}\fi

\bibitem[{{Barkana} \& {Loeb}(1999)}]{1999ApJ...523...54B}
{Barkana}, R., \& {Loeb}, A. 1999, \apj, 523, 54

\bibitem[{{Behroozi} {et~al.}(2013){Behroozi}, {Wechsler}, \&
  {Conroy}}]{2013ApJ...770...57B}
{Behroozi}, P.~S., {Wechsler}, R.~H., \& {Conroy}, C. 2013, \apj, 770, 57

\bibitem[{{Boylan-Kolchin} {et~al.}(2011){Boylan-Kolchin}, {Bullock}, \&
  {Kaplinghat}}]{2011MNRAS.415L..40B}
{Boylan-Kolchin}, M., {Bullock}, J.~S., \& {Kaplinghat}, M. 2011, \mnras, 415,
  L40

\bibitem[{{Boylan-Kolchin} {et~al.}(2012){Boylan-Kolchin}, {Bullock}, \&
  {Kaplinghat}}]{2012MNRAS.422.1203B}
---. 2012, \mnras, 422, 1203

\bibitem[{{Brook} \& {Di Cintio}(2015)}]{2015MNRAS.450.3920B}
{Brook}, C.~B., \& {Di Cintio}, A. 2015, \mnras, 450, 3920

\bibitem[{{Brooks} {et~al.}(2013){Brooks}, {Kuhlen}, {Zolotov}, \&
  {Hooper}}]{2013ApJ...765...22B}
{Brooks}, A.~M., {Kuhlen}, M., {Zolotov}, A., \& {Hooper}, D. 2013, \apj, 765,
  22

\bibitem[{{Brooks} \& {Zolotov}(2014)}]{2014ApJ...786...87B}
{Brooks}, A.~M., \& {Zolotov}, A. 2014, \apj, 786, 87

\bibitem[{{Bryan} \& {Norman}(1998)}]{1998ApJ...495...80B}
{Bryan}, G.~L., \& {Norman}, M.~L. 1998, \apj, 495, 80

\bibitem[{{Bullock} {et~al.}(2000){Bullock}, {Kravtsov}, \&
  {Weinberg}}]{2000ApJ...539..517B}
{Bullock}, J.~S., {Kravtsov}, A.~V., \& {Weinberg}, D.~H. 2000, \apj, 539, 517

\bibitem[{{Chan} {et~al.}(2015){Chan}, {Kere{\v s}}, {O{\~n}orbe}, {Hopkins},
  {Muratov}, {Faucher-Gigu{\`e}re}, \& {Quataert}}]{2015MNRAS.454.2981C}
{Chan}, T.~K., {Kere{\v s}}, D., {O{\~n}orbe}, J., {et~al.} 2015, \mnras, 454,
  2981

\bibitem[{{Del Popolo} \& {Le Delliou}(2014)}]{2014JCAP...12..051D}
{Del Popolo}, A., \& {Le Delliou}, M. 2014, \jcap, 12, 051

\bibitem[{{Dutton} {et~al.}(2016){Dutton}, {Macci{\`o}}, {Frings}, {Wang},
  {Stinson}, {Penzo}, \& {Kang}}]{2016MNRAS.457L..74D}
{Dutton}, A.~A., {Macci{\`o}}, A.~V., {Frings}, J., {et~al.} 2016, \mnras, 457,
  L74

\bibitem[{{Elbert} {et~al.}(2015){Elbert}, {Bullock}, {Garrison-Kimmel},
  {Rocha}, {O{\~n}orbe}, \& {Peter}}]{2015MNRAS.453...29E}
{Elbert}, O.~D., {Bullock}, J.~S., {Garrison-Kimmel}, S., {et~al.} 2015,
  \mnras, 453, 29

\bibitem[{{Garrison-Kimmel} {et~al.}(2014){Garrison-Kimmel}, {Boylan-Kolchin},
  {Bullock}, \& {Kirby}}]{2014MNRAS.444..222G}
{Garrison-Kimmel}, S., {Boylan-Kolchin}, M., {Bullock}, J.~S., \& {Kirby},
  E.~N. 2014, \mnras, 444, 222

\bibitem[{{Governato} {et~al.}(2010){Governato}, {Brook}, {Mayer}, {Brooks},
  {Rhee}, {Wadsley}, {Jonsson}, {Willman}, {Stinson}, {Quinn}, \&
  {Madau}}]{2010Natur.463..203G}
{Governato}, F., {Brook}, C., {Mayer}, L., {et~al.} 2010, \nat, 463, 203

\bibitem[{{Guo} {et~al.}(2010){Guo}, {White}, {Li}, \&
  {Boylan-Kolchin}}]{2010MNRAS.404.1111G}
{Guo}, Q., {White}, S., {Li}, C., \& {Boylan-Kolchin}, M. 2010, \mnras, 404,
  1111

\bibitem[{{Jetley} {et~al.}(2010){Jetley}, F., {Mendes}, {Kale}, \&
  {Quinn}}]{changa2010}
{Jetley}, P., F., G., {Mendes}, C., {Kale}, L., \& {Quinn}, T. 2010, IEEE
  International Conference for High Performance Computing, Networking, Storage
  and Analysis (SC), 1

\bibitem[{{Jiang} \& {van den Bosch}(2015)}]{2015MNRAS.453.3575J}
{Jiang}, F., \& {van den Bosch}, F.~C. 2015, \mnras, 453, 3575

\bibitem[{{Kazantzidis} {et~al.}(2004){Kazantzidis}, {Mayer}, {Mastropietro},
  {Diemand}, {Stadel}, \& {Moore}}]{2004ApJ...608..663K}
{Kazantzidis}, S., {Mayer}, L., {Mastropietro}, C., {et~al.} 2004, \apj, 608,
  663

\bibitem[{{Klypin} {et~al.}(1999){Klypin}, {Kravtsov}, {Valenzuela}, \&
  {Prada}}]{1999ApJ...522...82K}
{Klypin}, A., {Kravtsov}, A.~V., {Valenzuela}, O., \& {Prada}, F. 1999, \apj,
  522, 82

\bibitem[{{Klypin} {et~al.}(2011){Klypin}, {Trujillo-Gomez}, \&
  {Primack}}]{2011ApJ...740..102K}
{Klypin}, A.~A., {Trujillo-Gomez}, S., \& {Primack}, J. 2011, \apj, 740, 102

\bibitem[{{Kravtsov} {et~al.}(2004){Kravtsov}, {Berlind}, {Wechsler}, {Klypin},
  {Gottl{\"o}ber}, {Allgood}, \& {Primack}}]{2004ApJ...609...35K}
{Kravtsov}, A.~V., {Berlind}, A.~A., {Wechsler}, R.~H., {et~al.} 2004, \apj,
  609, 35

\bibitem[{{Kuijken} \& {Dubinski}(1995)}]{1995MNRAS.277.1341K}
{Kuijken}, K., \& {Dubinski}, J. 1995, \mnras, 277, 1341

\bibitem[{{Martinez}(2015)}]{2015MNRAS.451.2524M}
{Martinez}, G.~D. 2015, \mnras, 451, 2524

\bibitem[{{Mastropietro} {et~al.}(2005){Mastropietro}, {Moore}, {Mayer},
  {Debattista}, {Piffaretti}, \& {Stadel}}]{2005MNRAS.364..607M}
{Mastropietro}, C., {Moore}, B., {Mayer}, L., {et~al.} 2005, \mnras, 364, 607

\bibitem[{{Mayer} {et~al.}(2001){Mayer}, {Governato}, {Colpi}, {Moore},
  {Quinn}, {Wadsley}, {Stadel}, \& {Lake}}]{2001ApJ...559..754M}
{Mayer}, L., {Governato}, F., {Colpi}, M., {et~al.} 2001, \apj, 559, 754

\bibitem[{{Mayer} {et~al.}(2007){Mayer}, {Kazantzidis}, {Mastropietro}, \&
  {Wadsley}}]{2007Natur.445..738M}
{Mayer}, L., {Kazantzidis}, S., {Mastropietro}, C., \& {Wadsley}, J. 2007,
  \nat, 445, 738

\bibitem[{{McConnachie}(2012)}]{2012AJ....144....4M}
{McConnachie}, A.~W. 2012, \aj, 144, 4

\bibitem[{{Menon} {et~al.}(2015){Menon}, {Wesolowski}, {Zheng}, {Jetley},
  {Kale}, {Quinn}, \& {Governato}}]{2015ComAC...2....1M}
{Menon}, H., {Wesolowski}, L., {Zheng}, G., {et~al.} 2015, Computational
  Astrophysics and Cosmology, 2, 1

\bibitem[{{Moster} {et~al.}(2010){Moster}, {Somerville}, {Maulbetsch}, {van den
  Bosch}, {Macci{\`o}}, {Naab}, \& {Oser}}]{2010ApJ...710..903M}
{Moster}, B.~P., {Somerville}, R.~S., {Maulbetsch}, C., {et~al.} 2010, \apj,
  710, 903

\bibitem[{{O{\~n}orbe} {et~al.}(2015){O{\~n}orbe}, {Boylan-Kolchin}, {Bullock},
  {Hopkins}, {Kere{\v s}}, {Faucher-Gigu{\`e}re}, {Quataert}, \&
  {Murray}}]{2015MNRAS.454.2092O}
{O{\~n}orbe}, J., {Boylan-Kolchin}, M., {Bullock}, J.~S., {et~al.} 2015,
  \mnras, 454, 2092

\bibitem[{{Ogiya} \& {Burkert}(2015)}]{2015MNRAS.446.2363O}
{Ogiya}, G., \& {Burkert}, A. 2015, \mnras, 446, 2363

\bibitem[{{Pillepich} {et~al.}(2014){Pillepich}, {Kuhlen}, {Guedes}, \&
  {Madau}}]{2014ApJ...784..161P}
{Pillepich}, A., {Kuhlen}, M., {Guedes}, J., \& {Madau}, P. 2014, \apj, 784,
  161

\bibitem[{{Quinn} {et~al.}(1996){Quinn}, {Katz}, \&
  {Efstathiou}}]{1996MNRAS.278L..49Q}
{Quinn}, T., {Katz}, N., \& {Efstathiou}, G. 1996, \mnras, 278, L49

\bibitem[{{Read} {et~al.}(2015){Read}, {Agertz}, \&
  {Collins}}]{2015arXiv150804143R}
{Read}, J.~I., {Agertz}, O., \& {Collins}, M.~L.~M. 2015, ArXiv e-prints,
  arXiv:1508.04143

\bibitem[{{Read} \& {Gilmore}(2005)}]{2005MNRAS.356..107R}
{Read}, J.~I., \& {Gilmore}, G. 2005, \mnras, 356, 107

\bibitem[{{Sawala} {et~al.}(2014){Sawala}, {Frenk}, {Fattahi}, {Navarro},
  {Bower}, {Crain}, {Dalla Vecchia}, {Furlong}, {Helly}, {Jenkins}, {Oman},
  {Schaller}, {Schaye}, {Theuns}, {Trayford}, \& {White}}]{2014arXiv1412.2748S}
{Sawala}, T., {Frenk}, C.~S., {Fattahi}, A., {et~al.} 2014, ArXiv e-prints,
  arXiv:1412.2748

\bibitem[{{Sawala} {et~al.}(2016){Sawala}, {Frenk}, {Fattahi}, {Navarro},
  {Bower}, {Crain}, {Dalla Vecchia}, {Furlong}, {Helly}, {Jenkins}, {Oman},
  {Schaller}, {Schaye}, {Theuns}, {Trayford}, \& {White}}]{2016MNRAS.457.1931S}
---. 2016, \mnras, 457, 1931

\bibitem[{{Shen} {et~al.}(2014){Shen}, {Madau}, {Conroy}, {Governato}, \&
  {Mayer}}]{2014ApJ...792...99S}
{Shen}, S., {Madau}, P., {Conroy}, C., {Governato}, F., \& {Mayer}, L. 2014,
  \apj, 792, 99

\bibitem[{{Springel} {et~al.}(2008){Springel}, {Wang}, {Vogelsberger},
  {Ludlow}, {Jenkins}, {Helmi}, {Navarro}, {Frenk}, \&
  {White}}]{2008MNRAS.391.1685S}
{Springel}, V., {Wang}, J., {Vogelsberger}, M., {et~al.} 2008, \mnras, 391,
  1685

\bibitem[{{Tollerud} {et~al.}(2014){Tollerud}, {Boylan-Kolchin}, \&
  {Bullock}}]{2014MNRAS.440.3511T}
{Tollerud}, E.~J., {Boylan-Kolchin}, M., \& {Bullock}, J.~S. 2014, \mnras, 440,
  3511

\bibitem[{{Tomozeiu} {et~al.}(2016){Tomozeiu}, {Mayer}, \&
  {Quinn}}]{2016ApJ...818..193T}
{Tomozeiu}, M., {Mayer}, L., \& {Quinn}, T. 2016, \apj, 818, 193

\bibitem[{{Wang} {et~al.}(2012){Wang}, {Frenk}, {Navarro}, {Gao}, \&
  {Sawala}}]{2012MNRAS.424.2715W}
{Wang}, J., {Frenk}, C.~S., {Navarro}, J.~F., {Gao}, L., \& {Sawala}, T. 2012,
  \mnras, 424, 2715

\bibitem[{{Weinberg} {et~al.}(2015){Weinberg}, {Bullock}, {Governato}, {Kuzio
  de Naray}, \& {Peter}}]{2015PNAS..11212249W}
{Weinberg}, D.~H., {Bullock}, J.~S., {Governato}, F., {Kuzio de Naray}, R., \&
  {Peter}, A.~H.~G. 2015, Proceedings of the National Academy of Science, 112,
  12249

\bibitem[{{Widrow} \& {Dubinski}(2005)}]{2005ApJ...631..838W}
{Widrow}, L.~M., \& {Dubinski}, J. 2005, \apj, 631, 838

\bibitem[{{Widrow} {et~al.}(2008){Widrow}, {Pym}, \&
  {Dubinski}}]{2008ApJ...679.1239W}
{Widrow}, L.~M., {Pym}, B., \& {Dubinski}, J. 2008, \apj, 679, 1239

\bibitem[{{Wolf} {et~al.}(2010){Wolf}, {Martinez}, {Bullock}, {Kaplinghat},
  {Geha}, {Mu{\~n}oz}, {Simon}, \& {Avedo}}]{2010MNRAS.406.1220W}
{Wolf}, J., {Martinez}, G.~D., {Bullock}, J.~S., {et~al.} 2010, \mnras, 406,
  1220

\bibitem[{{Zolotov} {et~al.}(2012){Zolotov}, {Brooks}, {Willman}, {Governato},
  {Pontzen}, {Christensen}, {Dekel}, {Quinn}, {Shen}, \&
  {Wadsley}}]{2012ApJ...761...71Z}
{Zolotov}, A., {Brooks}, A.~M., {Willman}, B., {et~al.} 2012, \apj, 761, 71

\end{thebibliography}

\end{document}